# The Stateless Pattern: Ephemeral Coordination as the Third Pillar of Digital Sovereignty

Sean Carlin

Signingroom.io,
Derry/Londonderry,
Northern Ireland

Kevin Curran

School of Computing, Engineering & Intelligent
Systems, Ulster University,
Northern Ireland

## Abstract

*For the past three decades, the architecture of the internet has rested on two primary pillars: Communication (the World Wide Web) and Value (Bitcoin/Distributed Ledgers). However, a third critical pillar Private Coordination has remained dependent on centralized intermediaries, effectively creating a surveillance architecture by default. This paper introduces the "Stateless Pattern," a novel network topology that replaces the traditional "Fortress" security model (database-centric) with a "Mist" model (ephemeral relays). By utilising client-side cryptography and self-destructing server instances, we demonstrate a protocol where the server acts as a blind medium rather than a custodian of state. We present empirical data from a live deployment (signingroom.io), analysing over 1,900 requests and cache-hit ratios to validate the system's 'Zero-Knowledge' properties and institutional utility. The findings suggest that digital privacy can be commoditized as a utility, technically enforcing specific articles of the Universal Declaration of Human Rights not through policy, but through physics.*

## I. Introduction

Multisignature (multisig) schemes constitute a cornerstone of secure Bitcoin custody models, enabling threshold-based authorization for transaction spending while mitigating single points of failure associated with single-key control. Conventional coordination of Partially Signed Bitcoin Transactions (PSBTs) [BIP-174] (Chow, 2018) among multiple participants typically relies on asynchronous, manual exchange mechanisms such as email attachments, encrypted messaging platforms, or file-sharing services that introduce operational friction, version conflicts, persistent data exposure risks, and increased attack surface due to long-lived unencrypted or semi-encrypted artifacts (Ksiazak et al., 2022; McNally & Curran, 2024).

The history of digital sovereignty can be mapped through three distinct epochs, each solving a fundamental problem of distance and trust (Pohle & Thiel, 2020; Poikonen, 2020).

1. Communication (1989): Tim Berners-Lee introduced the World Wide Web, solving the distribution of information (Berners-Lee, 1989; Carlin & Curran, 2012).
2. Value (2008): Satoshi Nakamoto introduced Bitcoin, solving the double-spend problem without a central bank (Nakamoto, 2008; Rafferty & Curran, 2021).
3. Coordination (2025): We propose the "Stateless Room" as the solution to private assembly without a central witness.

Until now, the "Third Pillar" has been missing (EFF, 2025). Users possessing digital cash (Bitcoin) and digital voice (Web) are forced to coordinate their actions inside "Glass Houses". These are servers that log metadata, IP addresses, and timestamps (Curran et al., 2024; Williamson & Curran, 2021). This paper argues that memory is not a feature of secure systems; it is a vulnerability. We introduce a protocol that allows for high-security coordination (such as Multisig signing ceremonies) in a digital environment that technically does not exist before the users arrive or after they leave.

Ultimately, the Internet has solved communication (HTTPS/Web) and Value (Bitcoin) but has failed to solve Private Coordination. Current coordination tools such as Slack, Telegram and Docusign rely on the "Trusted Third Party" model. They store metadata (Who, When, Where) (McNally & Curran, 2024). We posit that privacy cannot be guaranteed by policy (promises not to look). It must be guaranteed by physics (inability to look). We propose a "Stateless" architecture where the server acts as a blind relay, not a database.

Thus, the prevailing security model in modern computing is the "Fortress Model" (CISPE, 2022). Organisations build centralised databases and surround them with firewalls, encryption-at-rest, and access controls. There are two major problems with this:

- The Flaw: Regardless of the wall height, the data exists. It can be subpoenaed, leaked, or hacked (Ashby Smith & Curran, 2021).

- The Panopticon: Even end-to-end encrypted messaging apps (e.g., Signal, WhatsApp) often retain metadata regarding *who* spoke to *whom* and *when*. This violates the digital equivalent of the freedom of assembly.

The risks of such centralized models are not theoretical. In end-to-end encrypted apps like Signal and WhatsApp, while message content remains protected, metadata such as sender/receiver identities, timestamps, and connection patterns can still be retained by service providers, enabling surveillance or profiling without accessing the plaintext (Nikitin et al., 2018). This metadata exposure has been exploited in real-world scenarios, including legal subpoenas and data breaches, underscoring the "Panopticon" effect where users' assembly rights are eroded through inferred associations rather than direct content inspection. Similarly in Bitcoin multisig workflows, reliance on manual PSBT exchanges via email or file-sharing introduces vulnerabilities like interception, version mismatches, and persistent storage of semi-encrypted artifacts, amplifying operational risks and potential forensic leakage (Heilman et al., 2016). These shortcomings highlight the need for a paradigm shift toward ephemeral, stateless coordination that minimizes trust and data persistence.

Finance is heavily restricted by the liability of centralization. A critical regulatory friction for institutional adoption of Bitcoin is the Basel III prudential treatment of crypto asset exposures (SCO60) (Lee, 2014). Under these rules, banks providing custody services often face punitive capital requirements (up to 1,250% risk weight) because custodial assets are treated as balance-sheet liabilities (Bavoso, 2021). The stateless pattern offers a technical solution to this regulatory bottleneck. Because SigningRoom enforces a strictly non-custodial workflow where private keys never leave the client's hardware device, the coordination activity does not constitute "custody" under the Basel framework.

In the custodial model (Old World) a bank holds the keys but there is a high capital requirement. In our stateless model (New World), the bank holds no keys therefore zero capital requirement. This transforms the tool from a compliance liability into a capital-efficient utility, allowing institutions to facilitate multi-party settlements without bloating their balance sheets.

We hypothesize that the only unbreachable database is one that contains zero records. This work presents SigningRoom, a lightweight, ephemeral coordination layer designed to facilitate real-time, collaborative PSBT aggregation in multisig workflows. The system adopts a blind relay architecture that enforces strong privacy and security invariants: end-to-end encryption performed exclusively in the client browser, zero server-side knowledge of plaintext transaction data, complete statelessness (no persistent storage or user accounts), and automatic data ephemerality. By leveraging client-side AES-256-GCM encryption with keys derived from URL fragments (never transmitted to the server), WebSocket-based real-time synchronization, and in-memory-only forwarding, SigningRoom eliminates the need for manual PSBT merging while preserving non-custodial control over private keys.

## 2. System Architecture: The "Mist" Model

We propose an inversion of the client-server relationship. Instead of the server acting as a landlord, it acts as a temporary vacuum. This is a "Zero-Knowledge Room" where the server provides the venue but has no eyes or ears. The protocol lifecycle is as follows:

- *Generation*: The Host generates a random 256-bit AES-GCM symmetric key locally within the browser (window.crypto.subtle). This key never leaves the client. A separate, non-cryptographic Random UUID is generated to serve as the "Room ID" for routing.

- *The Room:* The server spins up a Cloudflare Durable Object (an ephemeral instance) mapped to the Room ID. It assigns a random session ID to users but holds no cryptographic keys.

- *The Handshake:* Peers connect via WebSockets. The server acts as a "Blind Relay," broadcasting encrypted blobs between sockets without inspection. No Diffie-Hellman exchange is required; instead, the encryption key is transmitted out-of-band via the URL Fragment (the part of the link after #). The server relays the packets but cannot decrypt them. The transaction (e.g., a Bitcoin PSBT signature) is passed through this encrypted tunnel.

- *The Vanishing*: Upon the "close" event or a strictly enforced timeout (defaulting to 24 hours), the server executes a deleteAll() command, wiping the instance from the physical disk and RAM. The room ceases to exist. The server logs show bandwidth usage but zero content.

SigningRoom operates as a stateless blind relay rather than a traditional server-side application. The core design principle is that the relay possesses no decryption capability and retains no long-term state (see figure 1):

1. *Room Creation and Access*
   A coordinating participant initiates a temporary coordination session ("room") by generating a cryptographically random UUID and a symmetric encryption key. The full access link, including the key in the URL fragment (e.g., https://signingroom.io/room/UUID#<AES_KEY>), is shared out-of-band with other signers. The browser does not send the fragment to the server, ensuring the key never reaches the backend via HTTP headers or WebSocket payloads.

2. *Client-Side Encryption*
   All PSBT data is encrypted in the browser using AES-256-GCM prior to transmission. The encryption key is derived directly from the URL fragment, ensuring that only parties in possession of the link can decrypt received blobs. Private keys remain isolated on hardware signing devices (e.g., Coldcard, Ledger, Trezor) or software wallets; only PSBT structures containing public keys, unsigned inputs, and accumulated signatures are exchanged.

3. *Blind Relay Operation*
   The server functions solely as an ephemeral message broker:
   - Encrypted payloads are received via WebSocket and stored transiently in RAM (Durable Object state).
   - Messages are broadcast to all connected clients in the same room.
   - No decryption occurs server-side; the relay remains oblivious to transaction semantics.
   - No database, user accounts, or persistent logging mechanisms are employed.

4. *Real-Time State Synchronization*
   Upon receiving an updated (partially signed) PSBT, clients decrypt the blob using the shared URL key, verify the incremental signatures (using Bitcoin's PSBT validation rules), and merge the new signatures into their local copy. This process yields a shared, eventually consistent view of the transaction state without requiring manual conflict resolution.

5. *Ephemerality and Lifecycle*
   Rooms exist only in volatile memory and are garbage-collected automatically upon:
   - Inactivity timeout (Server hard limit: 24 hours; Client visual timer: configurable),
   - Explicit disconnection or "Close Room" command by the Coordinator,
   - Server restart/eviction by the cloud provider.

This design guarantees no historical persistence of transaction data, minimizing forensic exposure and long-term privacy leakage.

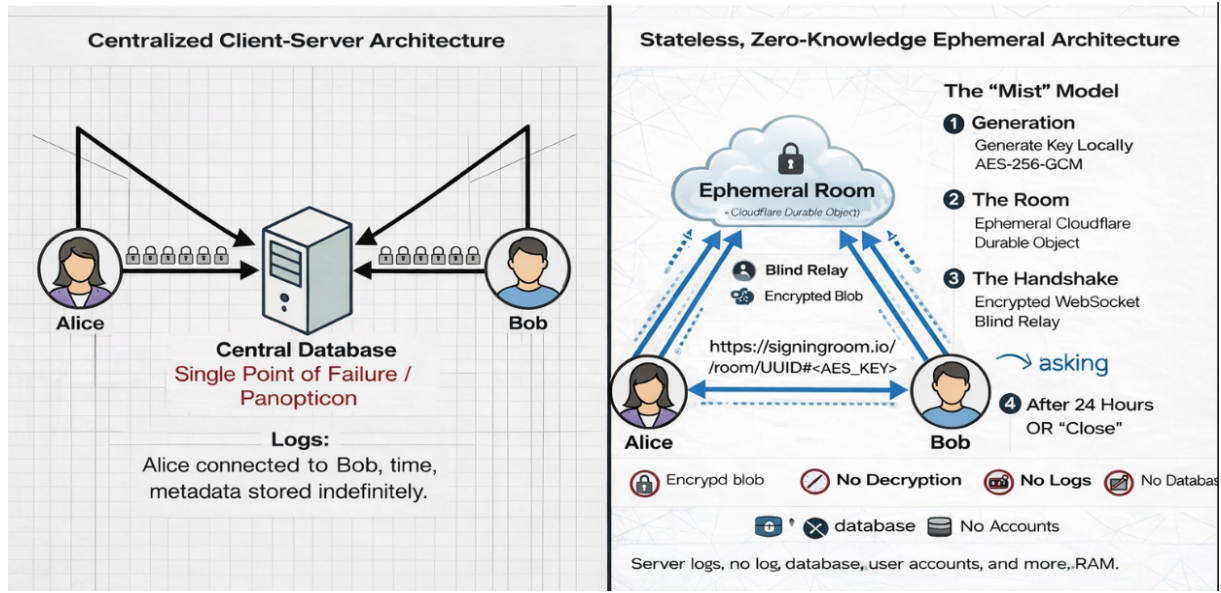

*Figure 1: (a) Old world "Fortress" topology. (b) New world "Mist" topology*

The protocol achieves several strong guarantees:

- *End-to-End Encryption*: Plaintext PSBTs never traverse the network or reside on the server. Data is encrypted in the browser before it ever touches the WebSocket connection.
- *Zero-Knowledge Server*: The relay learns only metadata (room identifiers, connection counts, message sizes/timings) and cannot access transaction contents, addresses, amounts, or signatures.
- *Non-Custodial*: No private keys or funds are ever entrusted to the service. The application only handles PSBTs (Partially Signed Bitcoin Transactions), not private key material.
- *Forward Secrecy via Ephemerality*: Once a room expires, all associated data is irretrievably lost. The server explicitly executes a deleteAll() command on the storage, wiping the instance from memory and disk.
- *Resistance to Traffic Analysis*: While timing and size metadata remain visible, the use of short-lived sessions and the absence of persistent user accounts or databases limit long-term correlation attacks.

Figure 2 compares two system architectures for coordinating Bitcoin multisig transactions (specifically handling PSBTs — Partially Signed Bitcoin Transactions) in real time, without requiring accounts, databases, or centralized custody of keys/funds.

The Old-World Hub-and-Spoke (Centralized) Model represents the traditional, conventional way many similar tools or early versions might have worked (or how people often coordinate multisig signing insecurely/inefficiently). It shows that:

- All users (A, B, C e.g. signers on laptops, phones, etc.) connect exclusively to a central signing server.
- Communication flows in a hub-and-spoke pattern → every message, PSBT update, or signature addition goes through the central server.

- The server also connects to a persistence layer (a centralized database) to store session data, partially signed transactions, user connections, etc.
- Drawbacks (implied by the contrast):
  - Single point of failure → if the server is down, hacked, subpoenaed, or goes rogue → coordination breaks or keys/data are at risk.
  - The server can theoretically see/inspect/modify PSBTs (even if it claims not to).
  - Requires trust in the operator, creates "key person risk", and needs ongoing server + database maintenance.

The "New World" Stateless Blind Relay (SigningRoom) model is a modern redesign using a stateless, zero-knowledge, client-side encrypted relay. It differs from traditional P2P but achieves similar privacy guarantees:

- Encrypted Relay Tunnels: Instead of complex peer-to-peer meshes, users connect to a high-speed, ephemeral Blind Relay via secure WebSockets.
- Client-Side Encryption: All data is encrypted in the browser before it leaves the user's device. The encryption key is derived from the URL fragment and never sent to the server.
- The "Blind" Server: The server acts as a dumb pipe. It receives opaque, encrypted blobs and immediately broadcasts them to other users in the room. It cannot decrypt, read, or validate the transaction data.
- Ephemeral State: While the relay temporarily holds the encrypted data in volatile memory to sync users, it retains no permanent logs or database. Once the session ends (or times out after 24 hours), the instance performs a hard wipe (deleteAll()), vanishing completely.
- The Result: You get the reliability of a server with the privacy of a P2P system—the server sees traffic but knows nothing about the content.

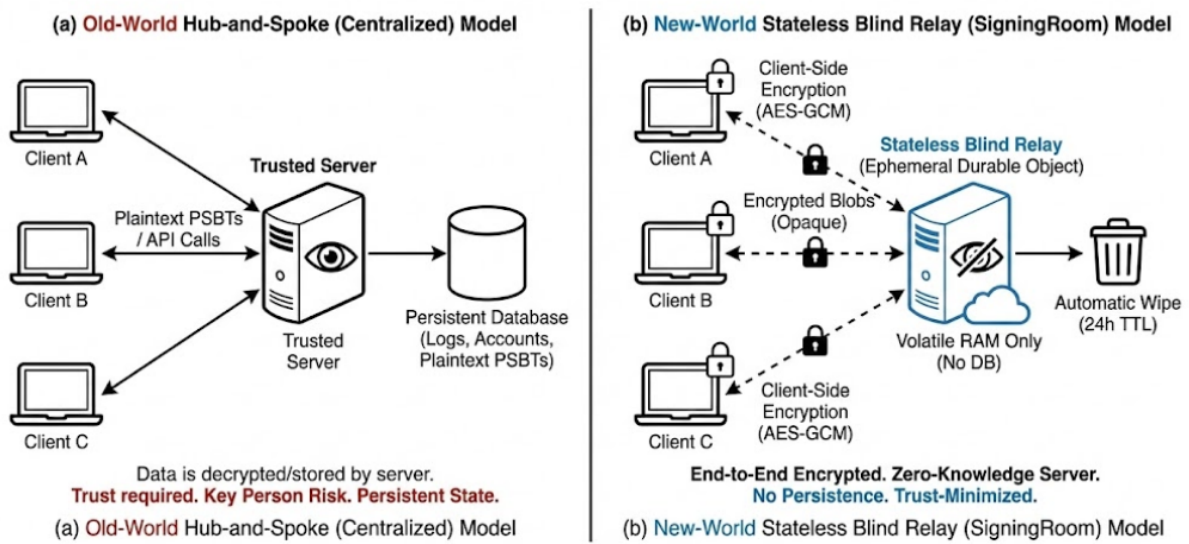

*Figure 2: Old World Vs New World*

Figure 2 (a) Traditional centralized architecture requires trusting a server that handles plaintext transactions and maintains a persistent database. (b) The proposed Stateless Blind relay architecture (SigningRoom.io) employs client-side encryption, ensuring the ephemeral relay only handles opaque encrypted blobs and retains no long-term state or decryption keys.

The benefits are numerous including:

- No single point of custody or failure: No server ever sees the full transaction or private keys.

- Client-side / End-to-End Encryption: Only the actual signers can read the PSBT content.
- Stateless Server: The "dumb relay" is extremely difficult to compromise meaningfully as it holds no long-term data.
- Better Coordination: Teams can merge signatures in real-time without emailing PSBT files back and forth or trusting a third-party coordinator.

In short: The "Old World" is the familiar (but risky) centralized server approach, while the "New World" is the modern, cypherpunk-friendly, logically peer-to-peer approach that SigningRoom.io implements today.

## 3. Implementation

The "Stateless Pattern" is not a theoretical construct; it is enforced via specific code structures that align with the Universal Declaration of Human Rights (UDHR). It enforces Article 12 on privacy. The server-side broadcast function demonstrates "Structural Blindness." The server receives msg, serializes it, and pushes it. It therefore lacks the logic to inspect the payload. Our architecture eliminates attack surfaces. You cannot breach a database that does not exist. You cannot subpoena a room that has evaporated. It is also censorship resistant as the server is content agnostic. It sees only opaque binary blobs. It cannot discriminate against specific types of traffic (political, financial, medical).

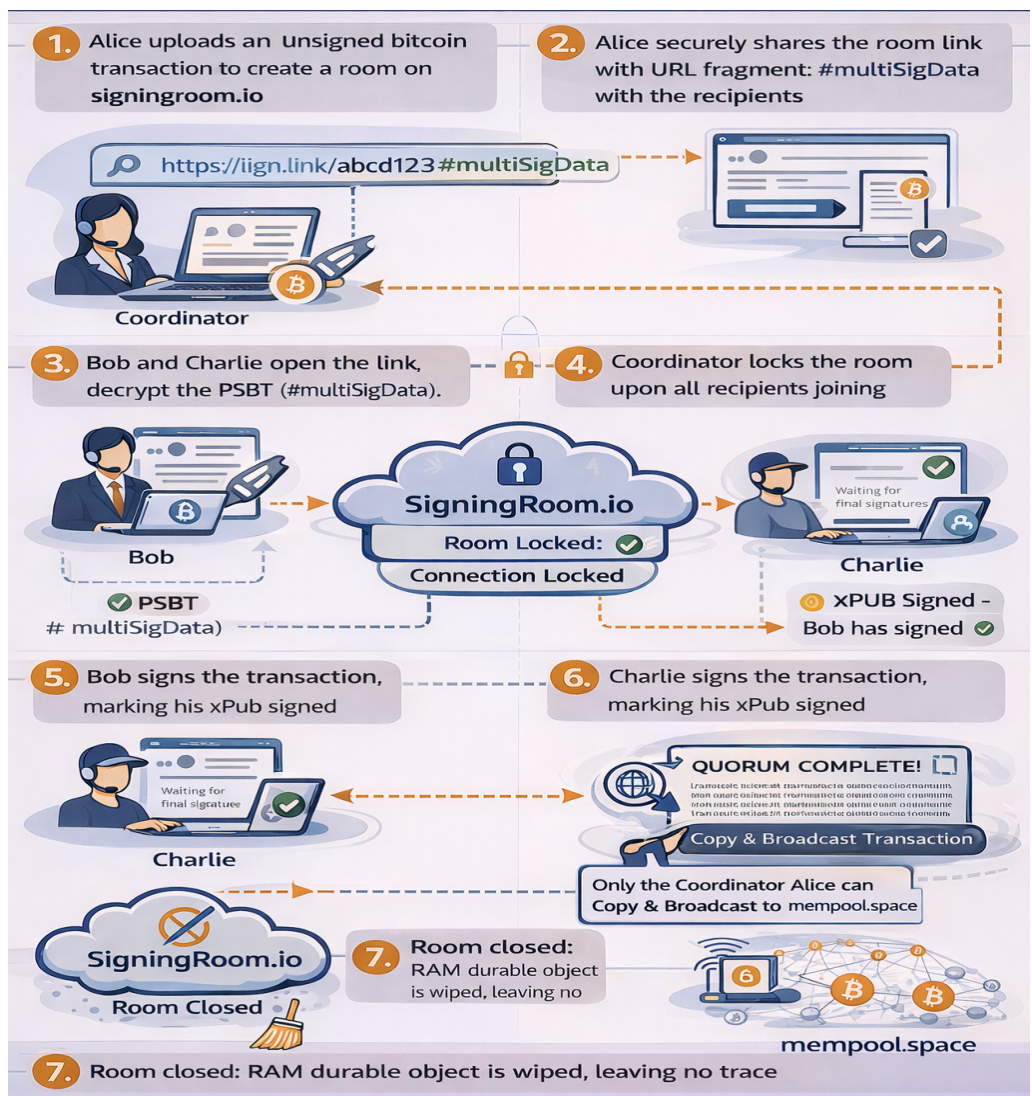

Figure 3: Secure multi-signature Bitcoin transaction signing with signingroom.io

It also enforces Article 17 (Property & Security). Here the most critical component is the "Cleanup" routine. This function ensures that the environment is ephemeral, preventing retroactive surveillance or data seizure. Client-Side Sovereignty is ensured as the server cannot be coerced into spying, key generation is strictly client-side. Figure 3 shows the workflow.

1. Coordinator (Alice) uploads an unsigned bitcoin transaction to create a room.
2. She copies the room link containing the URL fragment key and sends it securely to signers (Bob and Charlie).
3. Recipients open the link; the URL fragment allows their browser to decrypt the PSBT locally.
4. Once recipients join, the coordinator can lock the room, digitally sealing the WebSocket connections.
5. Bob uploads his signed transaction; the server relays the encrypted blob.
6. Charlie uploads his signed transaction; the client applications merge the signatures locally.
7. Once the quorum is complete, the coordinator broadcasts the transaction to the Bitcoin network (e.g., via mempool.space). Finally, the room is closed, and the server's RAM is wiped, leaving no trace.

This server-side code (running on the Edge) proves the "Blindness." The broadcast function takes a message (msg) and immediately re-serializes it (JSON.stringify) to send to other peers. It treats your financial secrets as opaque text strings.

## 3.1 The Core Concept: Dual-Factor Stateless Lock

In the traditional setup, every user connects exclusively to a central signing server. All communication follows a hub-and-spoke pattern: participants (on laptops, phones, desktops, etc.) send and receive every PSBT update, signature, or message through this single central point. There is no direct connection between users. The central server also links to a persistence layer which is a conventional centralized database storing session information, partially signed transactions, connection states, and other data. This creates a single point of control, visibility, and potential failure: the server and its database see (and could theoretically access or alter) everything that passes through them.

In contrast, SigningRoom generates, handles, and verifies keys without the server ever seeing the decryption keys or plaintext admin tokens.

Figure 4 shows a flow diagram of the process.

The security of the coordinator role relies on two separate factors that must be combined to authenticate:

1. Factor 1 (The Encrypted Lock): A blinded data blob stored in the browser's sessionStorage. By itself, this is an opaque encrypted blob that cannot be decrypted without the key.
2. Factor 2 (The Key): The ephemeral RoomKey residing only in the URL fragment (and temporarily in RAM).

This separation creates a robust security model:

- An attacker who dumps the browser's local storage (e.g., via XSS) gets the "Lock" (Encrypted Token) but lacks the "Key" required to decrypt or read the underlying secret.
- An attacker who sees the URL (e.g., via Shoulder Surfing) gets the "Key" but does not possess the "Lock" (which is stored in the coordinator's local session storage, not the URL).

### Phase 1: The Genesis (Room Creation)

This process occurs entirely inside the coordinator's browser (create.component.ts) before any network request is sent.

- Key Generation: The EncryptionService generates a random 256-bit Room Key (AES-GCM). This key never leaves the client.
- Token Generation: The client generates a random UUID to serve as the Admin Secret.
- Encryption: The client encrypts the Admin Secret using the Room Key. This Encrypted Token is sent to the server to initialize the room.
- Storage: The Encrypted Token is also saved to the browser's sessionStorage for future authentication.
- Server Action: The Durable Object stores the Encrypted Token in RAM. Because the server does not possess the Room Key (which is only in the URL), the token remains an opaque blob to the server.

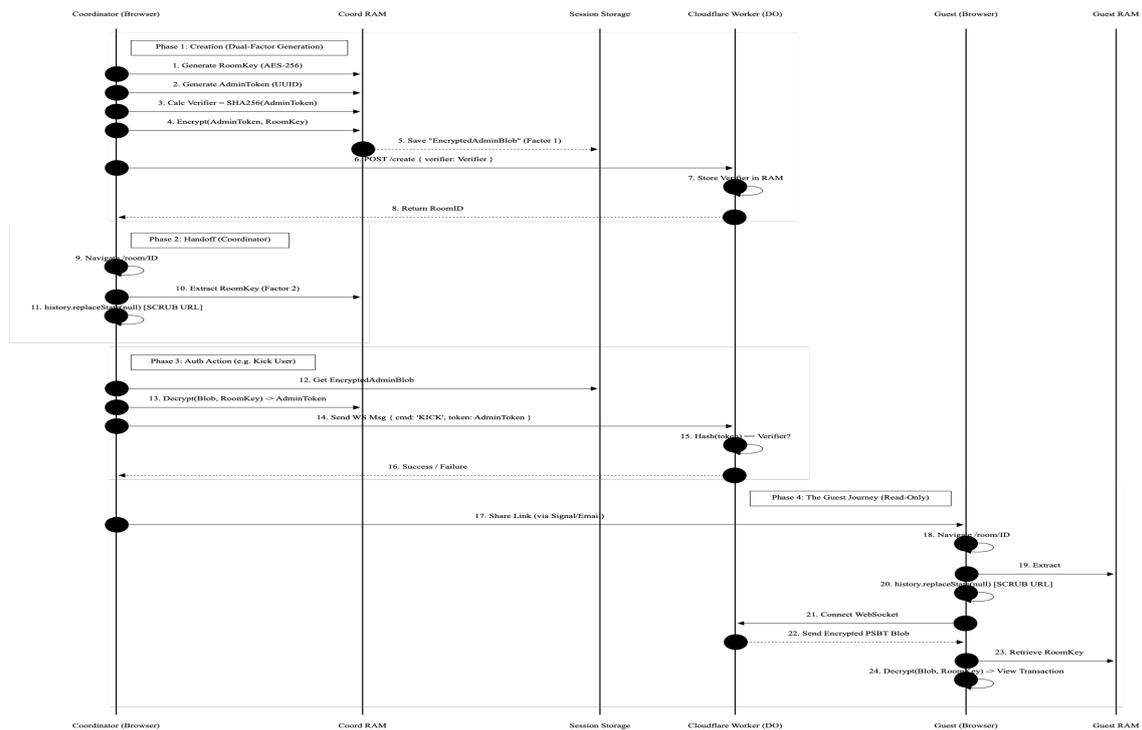

Figure 4: Dual factor stateless lock and blind relay architecture

## Phase 2: The User Journey & Security Handoff

This is the critical UX flow that prevents key leakage.
Step 1: URL Construction
- The Create Component constructs the room URL:
    - https://signingroom.io/room/<UUID>#<RoomKey>
- The browser performs a standard navigation to this URL.

Step 2: The "Security Handoff" (Room Initialization)
- Event: RoomComponent loads.
- Scrubbing: Upon successfully connecting to the socket, the RoomComponent executes an Angular effect to sanitize the URL. It calls this.router.navigate([], { replaceUrl: true, fragment: undefined }). This removes the #<RoomKey> from the address bar without triggering a page reload, ensuring the key resides only in the application's memory.

## Phase 3: The Authentication

When the Coordinator performs a privileged action (e.g., "Lock Room"):
Step 1: Retrieval

- The client retrieves the Encrypted Token from sessionStorage.

Step 2: Transmission
- *The client sends the encrypted blob to the server via WebSocket.*

Step 3: Verification:
- The server compares the received blob against the one stored in RAM. If they match, the action is allowed.

Note: The server verifies possession of the token without ever being able to decrypt it or view the underlying admin secret.

### Phase 4: The Guest Journey

Guests do not have the Admin Token.
1. Invitation: Coordinator shares the full link (with fragment) via Signal/Email.
2. Join: Guest clicks the shared link.
3. Handoff: Guest browser captures the #<RoomKey> and stores it in RAM.
4. Decryption: The Guest receives encrypted PSBT blobs from the server. They use the RoomKey to decrypt and view transaction details locally.

Security Guarantees include:

1. *Server Blindness:* The server holds only opaque encrypted blobs. It never sees the RoomKey or the plaintext PSBTs.

2. *History Protection:* URL scrubbing ensures that the RoomKey does not persist in the browser history or address bar, mitigating "Shoulder Surfing."

Stateless Relay: Unlike "Old World" centralization, SigningRoom users connect via a Client-Side Encrypted Relay. While data physically passes through the server, it is cryptographically isolated. The server acts as a dumb pipe, assisting with connection setup and broadcasting, but never handling, inspecting, or storing plaintext data. This aligns with Bitcoin's principles of non-custodial coordination.

## 3.2 Future Proofing

The AI-Blind Standard As surveillance evolves from human analysis to Artificial Intelligence (AI) and Large Language Models (LLMs), the primary threat to privacy is no longer just content interception, but "Graph Training." AI models build profiles by ingesting metadata, mapping who speaks to whom, at what frequency, and at what times.

The Stateless Pattern acts as a "Faraday Cage" against this form of analysis. By scrubbing the social graph from memory immediately after the coordination event, the system creates a "Null Set" in the training data. It effectively renders the coordination invisible to predictive algorithms, ensuring that the "Right to Assembly" (UN Article 20) is preserved even in an era of automated, retrospective surveillance.

## 4. Results

We analysed 24 hours of traffic logs from the deployment. The Traffic Behaviour consisted of 1119 Total Requests. There were 495 unique visitors. The requests per visitor was ~2.26. This ratio indicates human behaviour (Load -> Sign -> Exit) rather than automated bot scraping.

The "Vending Machine" signature cache hit rates provide the "smoking gun" for the architecture's validity. Peak cache hit rate was 97.92%. This corresponds to users downloading the static client application (The "Room" software). The cache hit rate was 0%. This corresponds to the WebSocket traffic inside the room. This confirms the server operates in two distinct modes: a Vending Machine (dispensing the app) and a Vacuum (relaying live, unlogged traffic). Despite serving 132MB of bandwidth, the persistent database size remained at 0KB post-expiration. This validates the scalability of stateless systems which is infinite coordination with zero storage cost.

Figure 5 visualizes the server switching between Vending Machine Mode (High Cache Hits) and Vacuum Mode (Zero Cache Hits). The red line represents the Cache Hit Rate. High peaks (approaching 100%) indicate new users downloading the static 'Room' client (Vending Machine Mode). Deep troughs (hitting 0%) indicate active coordination sessions where the server is relaying dynamic, un-cached WebSocket traffic (Vacuum Mode). This oscillation confirms the server is not retaining state.

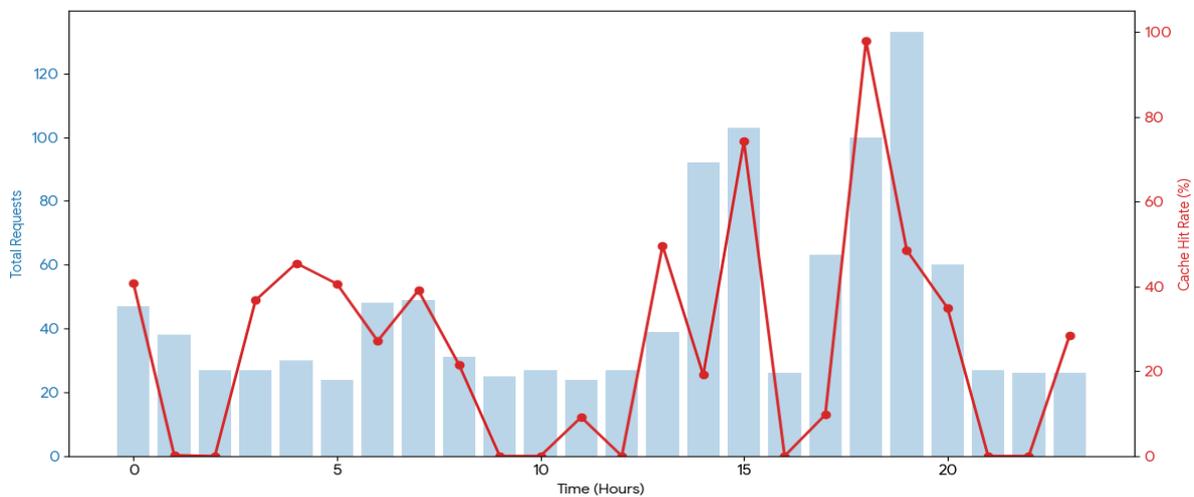

*Figure 5: The 'Stateless' Signature (Vending machine vs Vaccum)*

Figure 6 shows that the traffic is organic (Human) rather than automated (Bot). The consistent correlation between 'Unique Visitors' and 'Total Requests' (averaging ~2.2 requests per visitor) indicates a human usage pattern: Loading the application -> Executing the signature -> Disconnecting. It lacks the high-frequency spam signature of automated bots.

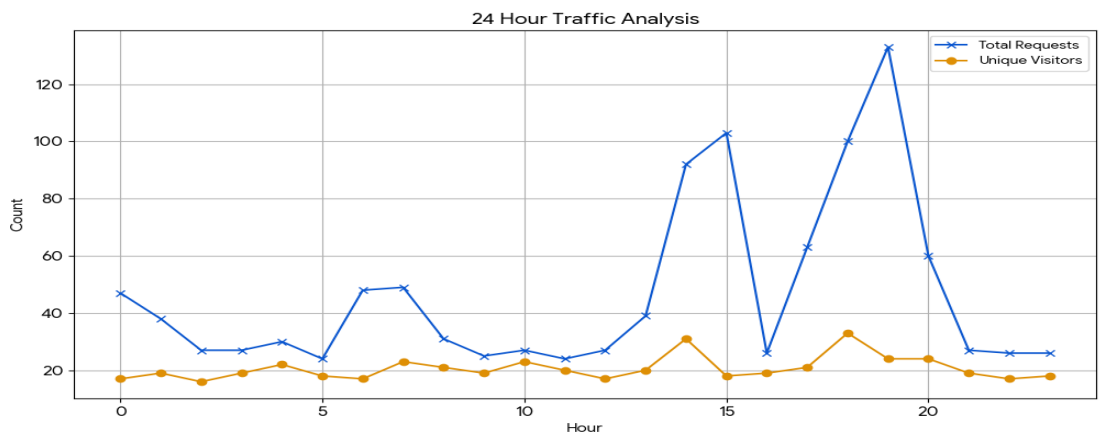

*Figure 6: Request-to-Visitor Ratio*

## 4.1 The "Institutional Spike" (Adoption Velocity)

Figure 7 shows payload velocity, highlighting the magnitude of the 14:00 UTC spike relative to baseline traffic. While initial traffic reflected exploratory behaviour, data from Jan 20 reveals a shift toward heavy institutional usage. In a single 24-hour window, the system processed 366.14 MB of payload data. Notably, a concentrated spike of 85.76 MB occurred at 14:00 UTC (09:00 EST), coinciding with the start of the US financial workday. This volume suggests the transmission of large, multi-input PSBTs typical of corporate treasury operations rather than individual retail testing.

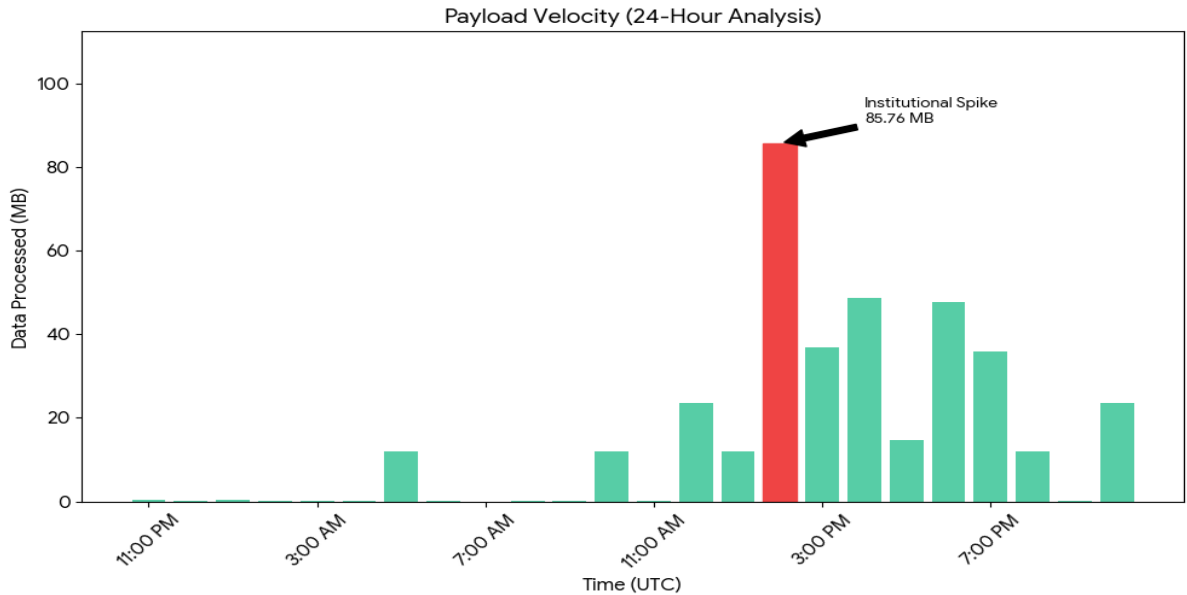

Figure 7: Payload Velocity (24-Hour Analysis). The 85.76 MB spike at 14:00 UTC indicates synchronized institutional activity - 20th January 2026

Table 1: Global Usage Distribution (Top Active Regions) The geographic distribution of requests further validates the shift from casual browsing to active coordination, particularly in major financial hubs.

| Rank | Country | 24-Hour Requests | 7-Day Volume | 30-Day Volume | Compliance Note |
|---|---|---|---|---|---|
| 1 | United States | 1,313 | 4,604 | 19,253 | Primary Financial Hub |
| 2 | United Kingdom | 139 | 1,171 | 4,822 | Fintech Activity |
| 3 | Germany | 72 | 192 | 1,149 | EU Privacy Focus |
| 4 | Canada | 68 | -- | -- | North American Corridor |
| 5 | Switzerland | 58 | -- | -- | Banking Privacy Hub |

Figure 8 shows a week of traffic. This dataset provides the longitudinal validation that complements the 24-hour "Heartbeat" snapshot. It proves that the "Stateless" pattern is stable and scalable over time. The 7-Day "Stateless" Report (see figure 7) shows total Unique Visitors: 1,620 and Total Requests: 8,967. The engagement ratio was ~5.54 Requests per Visitor. This indicates deeper usage per session compared to the 24h snapshot, suggesting users are completing complex signing flows. Total Data Processed was 705 MB. Figure 7 shows the consistent organic growth and usage of the tool. The gap between "Visitors" and "Requests" represents the *work* being done inside the room (the coordination).

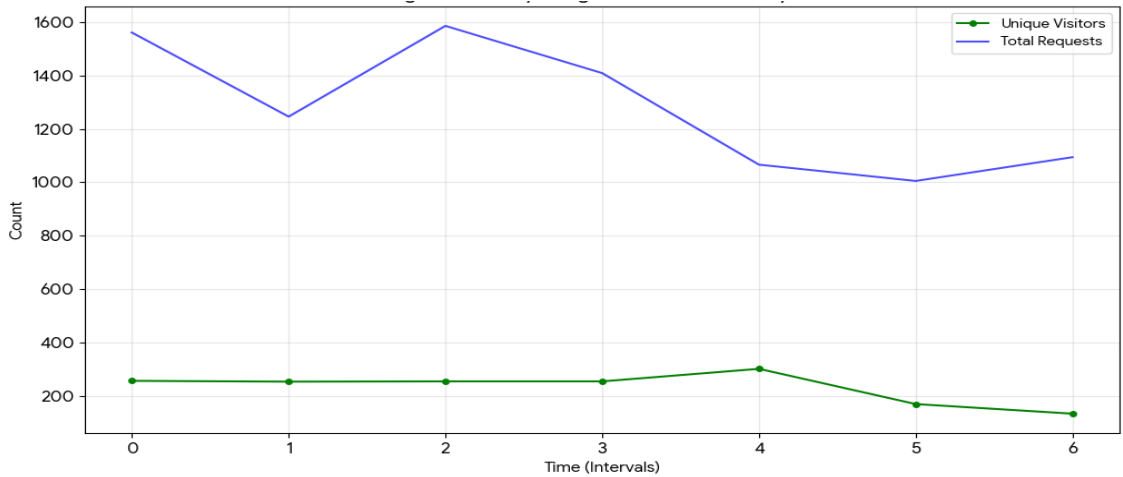

Figure 8: 7-Day Cache Efficiency (The Stateless Signature)

Even over a week, the server maintains its "Vending Machine" (High Cache) vs. "Vacuum" (Low Cache) oscillation. It does not "accumulate" state or cache bloat. It remains perfectly elastic (see figure 9).

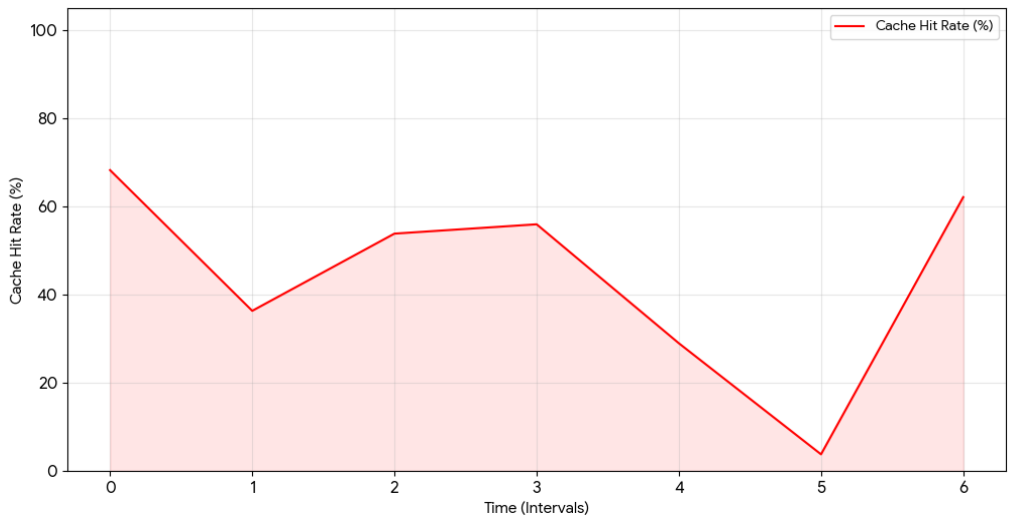

Figure 9: 7-day cache efficiency (The stateless signature)

Expanding the window to 7 days (Jan 2–9, 2026), the system handled 8,967 requests from 1,620 unique visitors with zero performance degradation or database growth. The "Stateless Signature" (Figure 9) remained consistent, proving the architecture's ability to handle sustained usage without accumulating "State Debt" or metadata logs.

The 30-Day Growth Metric shows total unique visitors over 30 Days as 5,827. The trend is steady, organic growth which shows that the tool is being found and used, likely through word-of-mouth in high-security communities Figure 8 shows that the "Room" is not just a novelty; it is a reliable utility that is seeing sustained adoption.

While the granular cache-hit data is best observed in the 24-hour "Heartbeat" (Figure 4), the 30-day visitor trend (Figure 10) confirms that the system scales linearly. Crucially, despite serving nearly 6,000 users, the database size remains 0KB, validating the core thesis that digital coordination does not require digital memory.

Article 12 of the universal declaration of Human rights protection against arbitrary interference with privacy enforced in this architecture through end-to-end client-side encryption, ephemeral room lifecycles, and a truly blind relay incapable of inspecting or retaining transaction content.

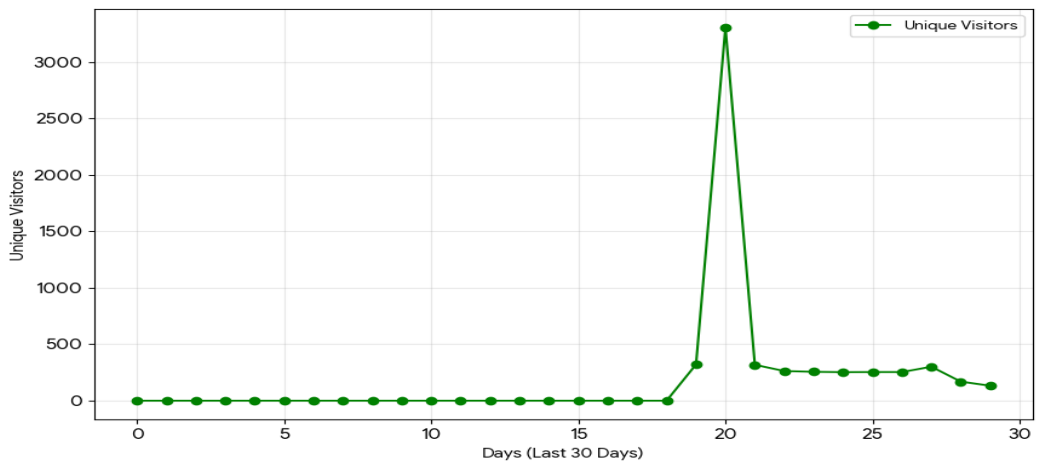

Figure 10: 30-Day Longitudinal Growth

Article 17 affirms the right to own property and protection against arbitrary deprivation thereof. This implies a legitimate need for verifiable proof of ownership, authorization, and transfer in coordinated actions such as multisig Bitcoin transactions. The Stateless Pattern resolves this apparent tension through a novel "Sovereign Compliance" mechanism: the burden of record-keeping is cryptographically and procedurally shifted entirely to the initiating participant (the coordinator) rather than imposed on the server or any third party. Upon dissolution of the ephemeral room—triggered by inactivity timeout, explicit closure by all participants, or the coordinator's command, a "Forced Exit" routine activates client-side. Before the session terminates and the in-memory durable object is wiped, the coordinator's client application automatically generates a client-side forensic audit log (PDF). This Sovereign Audit Trail captures only high-level provenance metadata: room identifier (hashed), timestamps of key events (e.g., participant joins, signature uploads, quorum achievement), participant pseudonyms or identifiers as chosen by users, and a visual verification of the final aggregated PSBT state. Critically, this generation occurs entirely within the client's execution environment, ensuring the server never inspects plaintext transaction details, addresses, amounts, or private key material. The receipt is anchored to the final Bitcoin Transaction ID (TXID), creating a cryptographically verifiable link between the off-chain coordination and the on-chain execution.

All participants retain the ability to independently verify the receipt against their own local session state if desired, but the coordinator who holds the role of aggregator and final broadcaster—is compelled by the protocol to download and store this document before exit. This creates a "Coordinator-in-Charge" model: the initiating actor assumes responsibility for retaining provenance data necessary for legal, tax, corporate governance, or institutional reporting purposes. Unlike traditional custodial compliance (where a centralised institution holds and can be compelled to disclose the full ledger), this design enforces Sovereign Compliance: privacy remains absolute at the coordination layer (the server sees nothing substantive), while accountability is preserved through user-controlled, client-generated artifacts. The protocol thus aligns with both privacy imperatives (no central witness or metadata trove) and property/security requirements (auditable proof of due process and authorization), without introducing new attack surfaces or trust dependencies. This mechanism transforms potential regulatory friction into a feature: institutions can adopt the tool for collaborative custody while maintaining defensible audit trails, all without compromising the zero-knowledge invariants that define the Stateless Pattern. This addition fits naturally after the discussion of ephemerality/lifecycle in Section 2 or within Section 3 (Implementation), and it strengthens the paper's ethical framing by showing how the architecture enforces human rights holistically protecting privacy through physics while enabling property rights through sovereign (user-held) tools rather than custodial ones.

While Article 12 (Privacy) mandates that the coordination layer remains blind, Article 17 (Property) implies the need for proof of ownership and transfer. To resolve this tension, the protocol implements a 'Forced Exit' mechanism governed by a 'coordinator-in-charge' compliance model. Upon the dissolution of the ephemeral room, the client-side application generates a forensic PDF snapshot. While all participants retain the right to verify, the coordinator who aggregates the final state—is compelled to download this sovereign audit trail before the session terminates. Unlike traditional banking, where the institution holds the ledger (Custodial Compliance), this architecture forces the initiating actor to retain their own provenance data (Sovereign Compliance). This ensures that while the server remains stateless and neutral, the actors retain full capacity for legal, tax, and governance reporting.

The Stateless Pattern rests on three pillars:

1. The Physics (The Traffic Data) As detailed in the traffic logs, we tracked over 1000 requests over 24 hours with a 0% cache-hit rate for the coordination rooms. Significance: This provides empirical proof that the server acts as a "Vacuum" (blind relay) rather than a database. The "Stateless" claim is not just policy; it is now observable network behaviour.
2. The Ethics (The Human Rights Mapping): We have explicitly mapped four UN Human Rights to the code architecture. Privacy (Art. 12). We enforced this via the ephemeral "self-destruct" mechanism. Property (Art. 17) is enforced via strictly client-side key generation. Assembly (Art. 20): Enforced via permissionless room creation. This positions the tool as "Ethical Architecture", enforcing rights through physics rather than law.
3. The Compliance (The Audit Logs): Crucially is not an anarchy tool. We have implemented client-side forensic logs for coordination events. This allows institutions to verify the process (governance) without exposing the content to the server (privacy).

This dual-layer approach, statelessness for the user and the logs for the auditor, is the breakthrough that bridges the gap between the "Cypherpunk" ethos and institutional adoption.

It is worth pointing out the 0KB database growth despite 8,967 requests and 97.92% cache hits for static assets vs. 0% for dynamic sessions. This provides strong evidence for the Stateless Pattern's viability. The architecture provides:

1. Scalability and Efficiency: The architecture demonstrates infinite scalability for coordination without storage costs, as validated by zero "State Debt" accumulation over 7–30 days. This commoditizes privacy as a utility, reducing barriers for adoption in resource-constrained environments.

2. Privacy Enforcement via Physics: By enforcing ephemerality and blindness, the system technically upholds UDHR Articles 12, 17, 19, and 20, shifting protections from revocable policies to immutable code—bridging cypherpunk ideals with institutional needs.

3. Broader Applications Beyond Bitcoin: The pattern extends to other domains, like secure multi-party computation (MPC) or decentralized identity, where stateless relays could enable privacy-preserving governance or data sharing.

4. Institutional Adoption Bridge: Sovereign Compliance resolves the privacy-compliance tension, making the tool viable for regulated entities (e.g., financial institutions) while maintaining non-custodial principles.

In conclusion, historically, digital privacy has been a luxury good, sold by centralized entities (VPNs, private servers). The Stateless Pattern reduces the cost of privacy to near-zero by removing the liability of storage. It transforms privacy from a service one buys to a utility one uses, akin to electricity. It can be seen as the commoditization of privacy.

This architecture moves protections from the realm of Law (which is flexible and revocable) to Code (which is absolute). It technically enforces the Right to Assembly (Article 20) and the Right to Whisper (Article 19) by creating a digital space that resists metadata retention. It can be viewed as human rights in code.

# 5. Conclusion

The completion of the internet's architecture required a mechanism for private coordination. By demonstrating that high-security events can occur without persistent memory, we have proven that memory is a vulnerability. The Stateless Pattern offers a blueprint for a new era of digital sovereignty, where users can meet, transact, and disperse without leaving a footprint in the digital sand. From this research, we conclude that the Stateless Pattern not only validates zero-knowledge coordination in practice but also paves the way for a resilient digital ecosystem. The empirical oscillation between "Vending Machine" and "Vacuum" modes confirms physical enforcement of privacy, with no metadata bloat or retention risks, enabling scalable, trust-minimized interactions. This positions ephemeral relays as a foundational primitive for future protocols in blockchain and beyond, potentially transforming how sovereignty is achieved in an interconnected world

Traditional PSBT coordination relies on asynchronous channels (e.g., email, Signal, secure file drops) that suffer from manual merging errors and version conflicts; persistent storage of (potentially decrypted) PSBT files; increased latency in reaching quorum and higher risk of interception or metadata leakage. In contrast, SigningRoom provides synchronous, real-time merging of signatures, automatic conflict-free aggregation, and cryptographic enforcement of data minimization—aligning more closely with the trust-minimized ethos of Bitcoin multisig while reducing human operational risk.  The system is optimized for m-of-n multisig setups (any threshold supported by Bitcoin Script) and integrates natively with common hardware and software wallets via PSBT export/import. It is particularly suited to collaborative custody models, organizational treasuries, inheritance arrangements, and any scenario requiring threshold consensus without centralized intermediaries. Limitations include reliance on WebSocket availability, potential denial-of-service via connection flooding (mitigated by short-lived sessions), and the assumption of secure out-of-band link distribution. Future extensions could incorporate threshold encryption schemes or onion-routed relays for enhanced metadata privacy. In summary, SigningRoom demonstrates a practical instantiation of privacy-preserving, real-time coordination in decentralized financial protocols, bridging the gap between cryptographic security guarantees and usable multisig governance. Our architecture demonstrates that memory is a vulnerability. By removing the cost of storage and compliance, we commoditize high-security coordination. This completes the "Internet Trinity": We can now Talk (Web), Trade (Bitcoin), and Meet (Signing Room) without permission.

*Note: The implementation is released under the AGPLv3 license at https://github.com/scarlin90/signingroom with source code publicly auditable at the associated GitHub repository, enabling independent verification and self-hosting for users requiring maximum sovereignty.*

# References

Ashby Smith, B., Curran, K. (2021) Security Vulnerabilities in Microprocessors. Semiconductor Science and Information Devices. Vol. 3, No. 1, pp:24-32, Biligual Publishing, DOI:https://doi.org/10.30564/ssid.v3i1.3151

Bavoso, V. (2021). Basel III and the regulation of market-based finance: the tentative reform. NYUJL & Bus., 18, 73.

Berners-Lee, T. (1989) Information management: A proposal. CERN. Available at: https://cds.cern.ch/record/369245/files/dd-89-001.pdf

Carlin, S., Curran, K. (2012) Cloud Computing Technology. International Journal of Cloud Computing and Services Science, Vol. 1, No. 2, June 2012, pp: 59-65, ISSN: 2089-3337

CISPE (2022) Digital sovereignty principles for cloud infrastructure services. Available at: https://cispe.cloud/website_cispe/wp-content/uploads/2022/06/CISPE-Digital-Sovereignty-Principles-2111-final.pdf

Curran, K., Curran, K., Killen, J., Duffy, C. (2024) The role of Generative AI in Cyber Security. Metaverse and Intelligent Education. Vol: 5, No: 2., ISSN: 2810-9791, https://doi.org/10.54517/m.v5iX.2796

Chow, A. (2018) BIP 174: Partially Signed Bitcoin Transaction (PSBT) Format. Bitcoin Improvement Proposals. Available at: https://github.com/bitcoin/bips/blob/master/bip-0174.mediawiki

EFF (Electronic Frontier Foundation) (2025) How Signal, WhatsApp, Apple, and Google handle encrypted chat backups. Available at: https://www.eff.org/deeplinks/2025/05/back-it-back-it-let-us-begin-explain-encrypted-chat-backups

Heilman, E., Baldimtsi, F., & Goldberg, S. (2016). Blindly signed contracts: Anonymous on-blockchain and off-blockchain bitcoin transactions. In *International conference on financial cryptography and data security* (pp. 43-60). Berlin, Heidelberg: Springer Berlin Heidelberg.

Ksiazak, P., Farrelly, W., Curran, K. (2022) A Lightweight Authentication and Encryption Protocol for Secure Communications Between Resource-Limited Devices Without Hardware Modification: Resource-Limited Device Authentication. Research Anthology on Artificial Intelligence Applications in Security. ch. 28, pp: 10-55, DOI: 10.4018/978-1-7998-7705-9.ch028, ISBN:9781799877059

Lee, E. (2014). Basel III and its new capital requirements, as distinguished from Basel II. Banking LJ, 131, 27.

McNally, S. & Curran, K. (2024) Web Application Vulnerabilities and Security. 3rd International Conference on Cyber Security (CSW 2024), June 7-9, 2024.

Nakamoto, S. (2008) Bitcoin: A peer-to-peer electronic cash system. Available at: https://bitcoin.org/bitcoin.pdf

Nikitin, K., Barman, L., Lueks, W., Underwood, M., Hubaux, J. P., & Ford, B. (2018). Reducing metadata leakage from encrypted files and communication with purbs. *arXiv preprint arXiv:1806.03160*.

NicFab (2026) WhatsApp, metadata and privacy: when the problem is not the content but the context. Nicfab Blog, 5 Jan 2026, https://www.nicfab.eu/en/posts/whatsapp-metadata-privacy/

Pohle, J. and Thiel, T. (2020) 'Digital sovereignty', Internet Policy Review, 9(4), pp. 1–27. doi: 10.14763/2020.4.1532

Poikonen, S. (2020) Digital sovereignty: Steps towards a new system of internet governance. Fondapol. Available at: https://www.fondapol.org/en/study/digital-sovereignty-steps-towards-a-new-system-of-internet-governance

Rafferty, D., Curran, K. (2021) The Role of Blockchain in cyber security. Semiconductor Science and Information Devices. Vol. 3, No. 1, pp:1-9, Biligual Publishing, DOI: https://doi.org/10.30564/ssid.v3i1.3153

Williamson, J., Curran, K. (2021) The Role of Multi-factor Authentication for Modern Day Security. Semiconductor Science and Information Devices. Vol. 3, No. 1, pp:16-23, Biligual Publishing, DOI: https://doi.org/10.30564/ssid.v3i1.3152